\documentclass[twocolumn,showpacs,amsmath,amssymb,prl,a4paper]{revtex4}

\newcommand{\Tr}{\mathop\mathrm{Tr}}

\newcommand{\expc}[1]{\left<#1\right>}
\newcommand{\re}{\mathop\mathrm{Re}}
\newcommand{\im}{\mathop\mathrm{Im}}

\newcommand{\clt}[1]{{#1}_t^{(c)}}
\begin{document}
\title{Quantum
dynamics and breakdown of classical realism in nonlinear oscillators}
\author{Omri Gat}
\affiliation{Racah Institute  of  Physics, Hebrew University
of Jerusalem, Jerusalem 91904, Israel}
\begin{abstract}
The dynamics of a quantum nonlinear oscillator is studied in terms of its quasi-flow, a dynamical mapping of the classical phase plane that 
represents the time-evolution of the quantum observables. Explicit expressions are derived for the deformation of the classical flow by the quantum nonlinearity in the semiclassical limit. The breakdown of the classical trajectories under the quantum nonlinear dynamics is quantified by the mismatch of the quasi-flow carried by different observables. It is shown that the failure of classical realism can give rise to a dynamical violation of Bell's inequalities. 
\end{abstract}
\pacs{\texttt{03.65.Sq, 42.50.Dv, 03.65.Fd}}
\maketitle
\paragraph{Introduction}
Single-degree of freedom nonlinear quantum oscillators are familiar as simple model systems and textbook examples of quantum mechanics. They are also of direct physical interest as effective Hamiltonians for the pardigm systems of trapped atoms and ions \cite{traps-dynamics,traps1,traps2} and single-mode cavity quantum electrodynamics and quantum optics \cite{imamoglu,qu-opt,cavity-qed}.

These systems are often studied experimentally by exciting them with an external perturbation and following the resultant evolution. The excited states are in general completely arbitrary: They are not necessarily close to specific energy states, nor to a superposition of a small number of energy states. For example, a nonlinear single-mode cavity can be excited into an arbitrary superposition of photon states, and a trapped atom may be excited into a state with an arbitrary wave function \cite{traps-dynamics}, and in principle these states can be tailored for specific purposes \cite{law-eberly}. At this stage it becomes important to develop a fundamental understanding and practical calculational tools for quantum nonlinear dynamics of oscillators of arbitrary form, with an arbitrary initial state. In this way, for example, it would be possible to predict the behavior of a nonlinear cavity as a light source.

In this Letter we put forward a method of general applicability for the study of the quantum dynamics of nonlinear oscillators, derived from the classical nonlinear dynamics. Since the classical dynamics is simple and well-understood, it serves as an excellent starting point for the analysis of the quantum dynamics. In this manner, the time evolution of quantum observables is studied here for oscillators of {arbitrary} form and strength of nonlinearity, with an arbitrary initial wavefunction. It is obtained as a power series in Planck's constant, where the semiclassical limit is approached when the level spacing changes slightly between one pair of levels and the next one, an assumption which is certainly justified in the applications cited above. 
On the other hand, it is \emph{not} assumed that the uncertainties are small, nor is any other assumption made on the properties of the quantum state.

The dynamics of the oscillator is studied via its \emph{quasi-flow}, a time-dependent mapping of the classical phase plane to itself that yields the dynamics of the observable for a given initial wavefunction. The quasi-flow is a smooth function of $\hbar$ which reduces to the classical flow when $\hbar\to0$. The analysis of the dynamics via the quasi-flow offers several advantages as compared with standard semiclassical methods. Unlike the linearization method \cite{haus-lai,lai-yu} no assumptions are made on the smallness of the uncertainties. We do not rely on the calculation of highly oscillatory integrals needed in the theory of semiclassical propagator \cite{gutzwiller,lj-wp}, and in this manner avoid caustic singularities. Furthermore, the analysis is exact in the nonlinearity, facilitating the explicit calculation of the quasi-flow for a large class of oscillators without recourse to numerical simulations, enabling the quantitative study of important fundamental effects arising from the quantum nonlinearity.

Several nonclassical effects are identified and analyzed in terms of the quasi-flow. Squeezing \cite{squeezing,haus-lai} in its general sense refers to the distortion of the uncertainties by the \emph{classical} nonlinearity, which sometimes produces sub-ground state fluctuations in one of the quadratures; in the present analysis squeezing is obtained almost trivially, since it occurs also for classical phase-space distribution undergoing nonlinear dynamics. A second effect is the {deformation} of the quasi-flow by the quantum nonlinear dynamics. In the semiclassical domain the quasi-flow deformation manifests as a drift in the mean values of observables. In particular, energy conservation combined with the noncommutativity of the observables implies a quantum drift in the {amplitude} of the oscillations.

Of course, quantum mechanics cannot be fully described in terms of a phase-space flow. This basic fact is reflected in the present analysis in that the quasi-flow is observable-specific. That is, each observable carries a quasi-flow modification, and the mismatch of quasi-flows is responsible for a fuzziness which spoils the classical realism that may be attributed to the initial state. We demonstrate this \emph{dynamical} breakdown of classical realism by calculating explicitly the quantum discrepancy for a generating function of observables, and showing that it can lead to a dynamical violation of Bell's inequalities.

Classical dynamics of integrable systems is most conveniently traced using action-angle variables. The dynamics in the original variables is then expressed as a composition of three canonical transformations. The quantum analog of the action-angle transformation is the unitary transformation to quantum normal form, developed in Refs.\ \cite{lj1,lj2}. Accordingly, we first calculate the quasi-flow of oscillators in normal form, which describes the dynamics of nonlinear oscillators up to unitary equivalence. Then we derive a composition formula for the phase-space representations of unitary transformations, thereby implicitly obtaining the quasi-flow for nonlinear oscillators in general form.

The nonlinear oscillators studied here are ideal system, where interactions with the environment and decoherence are neglected. These important effects lend themselves naturally to the dynamical approach taken here, and their study is planned in future works

\paragraph{The Weyl quasi-flow}
As is well known, although position and momentum are not simultaneously measurable in quantum mechanics, it is still possible to define phase-space quasi-probabilites \cite{mock}. The quasi-probabilities have to be accompanied by a symbol map, which assigns a phase-space function to each Hilbert space operator, such that the expectation value of the operator is equal to the phase space integral of the symbol with the quasi-probability as a measure.

The quasi-flow is the time-dependent mapping $z_t(z)$ of phase space to itself defined by the symbols $\bigl(q_t(z),p_t(z)\bigr)$ of the canonical position $\hat q_t$ and momentum $\hat p_t$ in the Heisenberg picture  (we use the symplectic notation $z^1=-z_2=q$, $z^2=z_1=p$ with implied summation over repeated indices). Since the symbols of $\hat q$ and $\hat p$ are $q(z)=q$ and $p(z)=p$ (respectively) the initial value of the quasi-flow is the identity mapping. The quasi-flow facilitates the calculation of expectation values of the canonical variables as a function of time for any initial state, and of any other observable by means of a {static} transformation, as demonstrated below.

Evidently, the form of the quasi-flow depends on the choice of the quasi-probability and symbol scheme. We choose here the Wigner-Weyl scheme \cite{mock}, since it has several convenient properties, among them the simple form of the semiclassical expansion of the symbol of the product (the Weyl-{Groenewold}-Moyal product rule \cite{lj-wp}). The quasi-flow may be defined for any value of $\hbar$, but it has a particular advantage of reducing to the classical flow in the limit $\hbar\to0$, enabling us to take full advantage of the simplicity of the classical flow.

\paragraph{The quasi-flow of nonlinear oscillators in normal form}
Nonlinear oscillators are defined here as (spinless) one-degree of freedom quantum dynamical systems governed by Hamiltonians whose Weyl symbol, the classical Hamiltonian $H(q,p)$, has a single nondegenerate phase space minimum and no other fixed points.
It has been demonstrated in Ref.\ \cite{lj1} that such Hamiltonians are unitarily equivalent to a normal form Hamiltonian which commutes with the harmonic oscillator Hamiltonian. That is, there exists a unitary operator $\hat U$ such that
\begin{equation}\begin{split}
\hat Q&=\hat U^\dagger \hat q \hat U,\qquad \hat P=\hat U^\dagger \hat p \hat U\ ,\\
\hat H&=f(\hat B)\ ,\qquad
\hat B=\frac{\hat Q^2+\hat P^2}{2}\ .\label{eq:norm}
\end{split}\end{equation}
The classical limit of this transformation is a Cartesian version of the canonical transformation to action-angle variables; in particular the classical limit of $B$, the symbol of $\hat B$, is the classical action. Like its classical analog, the quantum normal form Hamiltonian is a representative of its equivalence class where dynamics takes a particularly simple form. For this reason we will assume here that the Hamiltonian has been reduced to normal form, construct its quasi-flow and study its properties, returning to the general case in the last part of the paper.

It is convenient to formulate the normal form dynamics in terms of the complex coordinate $\beta=\frac{1}{\sqrt2}(Q+iP)$ and the annihilation operator $b=\frac{1}{\sqrt 2}(\hat Q+i \hat P)$ (note that $b$ differs by a factor of $\sqrt\hbar$ from the usual definition). The symbol of $b_t$ will be denoted by $\beta_t$. In complex coordinates $B=|\beta|^2$ and $\hat B=\frac12(bb^\dagger+b^\dagger b)$.
 
Using the algebraic properties of the observables \cite{arik}, their dynamics in normal form can be expressed as a rotation with an operator-valued frequency
\begin{equation}\label{eq:qdynamics}
b_t\equiv e^{-\frac{\hat H}{i\hbar}t}be^{\frac{\hat H}{i\hbar}t}
=e^{-i\Omega_+(\hat B)t}b=b e^{-i\Omega_-(\hat B)t}\ ,
\end{equation}
where $\Omega_{\pm}(\hat B)=\pm\frac{1}{\hbar}(f(\hat B\pm\hbar)-f(\hat B))$. Note that $\lim\limits_{\hbar\to0}\Omega_{\pm}=\omega$ is the classical (action-dependent) frequency.

Either of the two expressions on the right-hand-side of the last equation forms a useful starting point for a semiclassical expansion, since their factors have symbols with smooth dependence on $\hbar$. This fact makes the calculation of the semiclassical expansion of the normal-form quasi-flow a straightforward application of the rules of the Moyal algebra; the resulting expression to second order in $\hbar$ is
\begin{equation}\label{eq:betat-beta}
\beta_t(\beta)=\clt{\beta}(1+\hbar^2\beta_t^{(2)}+O(\hbar^4))\ ,
\end{equation}
where 
$
\clt{\beta}=\beta e^{-i\omega(B)t}
$
is the classical flow, and 
\begin{align}
\re\beta_t^{(2)}&=\tfrac14((\omega'(B)t)^2+B\omega'(B)\omega''(B)t^2)\ ,\label{eq:beta2}\\
\im\beta_t^{(2)}&=\tfrac{1}{24}(-2B(\omega'(B)t)^3+5\omega''(B)t+2B\omega'''(B)t)\ ,\label{eq:phi2}
\end{align}
are the leading quantum correction to the modulus and phase of $\beta_t$, respectively.

The physical significance of Eqs. (\ref{eq:betat-beta}--\ref{eq:phi2}) is as follows: The leading term describes the squeezing of the moments of $b$ by the phase-plane shearing implied by the action-dependence of the classical frequency. Since the leading term in the dynamics of all observables is given by the classical flow, this effect is equivalent to the drift and deformation of the initial (Wigner) quasi-distribution by the classical flow. The higher order terms, in contrast, express a {nonlinear quantum drift}, which has no classical analog. 
The imaginary part of the quantum drift generates an accelerating drift in the phase of $\expc{b_t}$, and the real part generates a drift in the {amplitude} of the oscillations; at first glance the existence of an amplitude drift seems surprising, since energy conservation implies conservation of the action operator $\hat B$. In fact, it is precisely the conservation of the action, which \emph{generates} the amplitude drift through the Moyal algebra.
Thus, on one hand $B_t(\beta)=B$, and on the other hand
\begin{multline}\label{eq:Bt}
B_t(\beta)=\frac{\beta_t\star\bar\beta_t+\bar\beta_t\star\beta}{2}=
\\\quad B +\hbar^2\bigl(2B\beta_t^{(2)}-\frac18\partial_\mu\partial\nu\clt{\beta}\partial^\mu\partial^\nu\clt{\bar\beta}\bigr)+O(\hbar^4)\ ,
\end{multline}
where $\star$ stand for the Moyal product, $\partial_\mu=\frac{\partial}{\partial z^\mu}$, and $\partial^\mu=-\frac{\partial}{\partial z_\mu}$. The two results imply that $\re\beta_t^{(2)}=\frac1{16B}\partial^\mu\partial^\nu\clt{\beta}\partial_\mu\partial_\nu\clt{\bar\beta}$, consistently with Eq.~(\ref{eq:beta2}).

\paragraph{Dynamical breakdown of classical realism}
Unlike the leading (classical) term of the quasi-flow, the higher order terms cannot be directly used to derive the dynamics of the quasi-distribution. That is, it is not possible to calculate the Wigner function of a state at time $t$, given the initial Wigner function and the quasi-flow Eqs. (\ref{eq:betat-beta}--\ref{eq:phi2}).

Thus, in particular, the quasi-flow in itself does not suffice for the calculation of higher moments of the canonical observables, such as the uncertainties. For this purpose the quasi-flow has to be supplemented by the \emph{discrepancy} $d_t(\hat A)$ of an observable $\hat A$ at time $t$, being the difference between the symbol of $\hat A_t$ and the symbol of $\hat A$ \emph{evaluated} at the quasi-flow $\beta_t$.  
The quantum discrepancy of a general observable in the $\hat q,\hat p$ algebra is extractable from the discrepancy of the generating function $e^{\alpha b^\dagger_t-\bar\alpha b_t}$ of the complex parameter $\alpha$,
\begin{multline}\label{eq:dy}
d_t(e^{\alpha b^\dagger_t-\bar\alpha b_t})
=\hbar^2e^{\alpha\clt{\bar\beta}-\bar\alpha \clt{\beta}}\\\times\bigl(P_1\omega'(B)t+P_2(\omega'(B)t)^2+P_3(\omega'(B)t)^3\quad\\+P_4\omega''(B)t+P_5\omega'(B)\omega''(B)t^2
\bigr)+O(\hbar^4)\ ,
\end{multline}
where $P_1,\ldots,P_5$ are explicit low-degree polynomials in $B$, $|\alpha|^2$, $\alpha\clt{\bar\beta}$ and $\bar\alpha\clt{\beta}$. Together with the quasi-flow, Eq. (\ref{eq:dy}) gives the full information on the quantum dynamics of the (normal-form) nonlinear oscillator to $O(\hbar^4)$.

The irreducibility of the dynamics to a phase-space flow, as expressed by the quantum discrepancy, is a consequence of the breakdown of classical realism under the quantum nonlinearity. That is, suppose that the initial state is such that $\hat q$ and $\hat p$ have a classical hidden variable underpinning; the underpinning could be provided, for example, by the initial Wigner function, if it is nonnegative on the entire phase plane. Then the hidden variable underpinning breaks down once the system starts to evolve, and in general fails for $\hat q_t$ and $\hat p_t$ for any positive $t$, \emph{unless} the dynamics is linear.

It was pointed out in Ref. \cite{technion} that the breakdown of the hidden variable underpinning can be detected when the classical symbol of the observable attains values which do not belong the spectrum of the observable as a Hilbert space operator. Following Ref. \cite{technion}, such observables will be termed \emph{improper}. When the hidden variable underpinning is provided by a nonnegative Wigner function, its classical symbol is the Weyl symbol. It is a property of the Weyl calculus that the symbol of a function $f(\hat S)$ of a linear combination $\hat S$ of $\hat Q$ and $\hat P$ is equal to $f(S)$, and thus such observables are proper. However, if $f$ is nonlinear, then the quantum discrepancy $d_t{f(\hat S)}$ is nonzero for $t>0$, and $f(\hat S_t)$ may become improper. That is, the quantum discrepancy is a direct measure of the breakdown of classical realism. As an example we may consider the observable $\hat C=\cos{\alpha \hat P}$ (with symbol $C$), whose spectrum is bounded between $-1$ and $1$. It follows from Eq.\ (\ref{eq:dy}) that (in real coordinates)
\begin{equation}\label{eq:ct}
C_t(Q,0)=1+\hbar^2\tfrac{\alpha^2}{8}(3\omega'(\tfrac{Q^2}2)+Q^2\omega''(\tfrac{Q^2}2))\omega'(\tfrac{Q^2}2)t^2+O(\hbar^4)\ ,
\end{equation}
which implies that for a large class of nonlinear oscillators, $\hat C_t$ is improper for all $t>0$, and the failure of classical realism increases quadratically in time.

An important further consequence of the breakdown of classical realism 
arises when the nonlinear oscillator is entangled with other degrees of freedom. The entanglement is assumed to result from the initial preparation, after which the nonlinear oscillator evolves purely under the single-degree of freedom Hamiltonian. If the entangled state, like the EPR state \cite{technion,bell}, has a nonnegative Wigner function, 
then initially observations of Bell operators constructed from functions of linear combination of $\hat Q$ and $\hat P$ are guaranteed to satisfy Bell's inequalities. At later times, however, measurements of a Bell operator constructed, for example, from $\cos(\alpha \hat Q_t)$ and $\cos(\alpha \hat P_t)$ are liable to lead to  violate Bell's inequalities, since these operators are improper for $t>0$. As demonstrated by Eq.\ (\ref{eq:ct}), this nonlinear dynamical Bell inequality violation increases quadratically in time. Further study of this topic is beyond the scope of the present work.
 
\paragraph{Dynamics of nonlinear oscillators in general form.}
Since every nonlinear oscillator is unitarily equivalent to a normal form oscillator \cite{lj1}, the results presented above yield the quasi-flow of any given nonlinear oscillators after it has been transformed to its normal form. Nevertheless it is of interest to express the result in terms of the {original} observables $\hat q,\hat p$ rather than the transformed observables $\hat Q,\hat P$. 

In the classical limit, the nonlinear oscillator flow can be expressed as a composition of three canonical transformations $(q,p)\longrightarrow(A,\phi)\longrightarrow(A,\phi+\omega(A)t)\longrightarrow(q_t,p_t)$, where the first and last arrows are the transformation to action-angle variables and its inverse (respectively), and the middle arrow is the time-evolution canonical transformation. Analogously the quantum evolution is the composition of three unitary transformations, $(\hat q,\hat p)\longrightarrow(\hat Q,\hat P)\longrightarrow(\hat Q_t,\hat P_t)\longrightarrow(\hat q_t,\hat p_t)$. Each unitary transformation can be represented by its Weyl symbol. The Weyl representation for the normal form transformation was derived in Ref. \cite{lj1}, while the representation for the time evolution is the quasi-flow given by Eqs. (\ref{eq:betat-beta}--\ref{eq:phi2}).

Unlike classical canonical transformations, however, the Weyl representation of the composition of the transformations is not equal to the composition of Weyl representations. Therefore, in order to obtain the general-form quasi-flow, a composition formula for Weyl symbols is needed. That is, given that the Weyl symbol of $\hat A$ with respect to the canonical set $\hat Q,\hat P$ is $A(z)$, and that the symbols of $\hat Q$ and $\hat P$ with respect to the set $\hat q,\hat p$ are $Z^\mu(z)$, to find the Weyl symbol $a(z)$ of $\hat A$ with respect to $\hat q,\hat p$. In the following we derive a semiclassical expansion of $a(z)$ to $O(\hbar^4)$, which, together with the previous results yields the quasi-flow $z_t^\mu(z)$ of nonlinear oscillators in general form.

The composition formula is derived directly from the definition of the Weyl symbol. By definition $a(z)=\Tr\hat A\Delta(z)$, where $\Delta(z)=\int \frac{dz'}{h}e^{\frac{i}{\hbar}(\hat z^\mu-z^\mu)z'_\mu}$ are the basis operators for the Weyl representation \cite{mock} with respect to the set $\hat z$, while $\hat A=\int\frac{dz}{h}A(z)\tilde\Delta(z)$, where $\tilde\Delta(z)$ are the basis operators with respect to $\hat Z$. Therefore
\begin{equation}\label{eq:a1}
a(z)=\int\frac{dz'}{h}A(z')\Tr\tilde\Delta(z')\Delta(z)=\int{dz'}A(z'){\delta(z';z)}\ ,
\end{equation}
\begin{widetext}
where $\delta(z';z)$, the symbol of $\tilde\Delta(z')$ with respect to $\hat z$ divided by $h$, can be calculated with the help of the Moyal calculus, giving
\begin{equation}
{\delta(z';z)}=
\bigg(1-\hbar^2\Big((\frac{1}{16}\partial_k\partial_lZ^\mu\partial^k\partial^lZ^\nu)\partial_\mu\partial_\nu-(\frac{1}{24}\partial_kZ^\lambda\partial^k\partial^lZ^\mu\partial_lZ^\nu)\partial_\lambda\partial_\mu\partial_\nu\Big)\bigg)\delta(z'-Z(z))+O(\hbar^4)\ .
\end{equation}
Using this result back in Eq.\ (\ref{eq:a1}) gives the composition rule
\begin{equation}
a(z)=\mathring{A}(z,Z(z))\ ,\qquad \mathring{A}(z,z')=\bigg(1-\hbar^2\Big((\frac{1}{16}\partial_k\partial_lZ^\mu\partial^k\partial^lZ^\nu)\partial_\mu\partial_\nu+(\frac{1}{24}\partial_kZ^\lambda\partial^k\partial^lZ^\mu\partial_lZ^\nu)\partial_\lambda\partial_\mu\partial_\nu\Big)\bigg)A(z')\ .
\end{equation}
\end{widetext}

\paragraph{Conclusions.} The phase-space analysis of the dynamics in the Heisenberg picture was shown to furnish a powerful tool for the study of a broad class of nonlinear quantum systems. Fundamental nonclassical dynamical effects, including the dynamical breakdown of classical realism were studied nonperturbatively. The main results are the explicit expressions for the quasi-flow and the quantum discrepancy in terms of the nonlinear frequency function $\omega(B)$, which hold all the dynamical information on the nonlinear oscillator and imply the nonclassical phenomena.

\paragraph{Acknowledgement} I have benefited from fruitful discussions with R. Littlejohn and R. Zeitak.

\end{document}